\documentclass[11pt,twoside]{article}
\usepackage{CAGN2019}
\usepackage{graphicx,natbib}

\usepackage[T1]{fontenc} 

\usepackage{latexsym}
\usepackage{verbatim}

\usepackage{ifpdf}  
\ifpdf  
      \DeclareGraphicsExtensions{.pdf,.png,.jpg}  
\else  
      \DeclareGraphicsExtensions{.eps}  
\fi 

\begin{document}

\vskip 1.0cm
\markboth{R.~Costero}{Multiplicity of the Orion Trapezium stars}
\pagestyle{myheadings}
%
%
\vspace*{0.5cm}
\parindent 0pt{Honour talk}


\vspace*{0.5cm}
\title{Multiplicity of the Orion Trapezium stars}

\author{R.~Costero$^1$}
\affil{$^1$Instituto de Astronomia, UNAM, A.P.\,70-264, 04510 Ciudad de M\'exico, M\'exico}

\begin{abstract}

Somewhat outside the topic of this conference, some preliminary results on the ongoing spectroscopic study of the six brightest Orion Trapezium stars is presented here. The main purpose of this work is to better understand the multiplicity and stability of each of these subsystems and the dynamical future of the group. So far the most interesting results reached are: 1) The orbit of the secondary star of the  eclipsing Component A (V1016 Ori) is highly inclined with respect to the equatorial plane of its primary star. 2) The also eclipsing binary BM\,Ori (Trapezium Component B) does have a tertiary member with period about 3.5 years, as proposed by \cite{VitriKloch04}, and is the same as the companion recently found by the \cite{GRAVITY+18}. 3) Component D is indeed a spectroscopic and interferometric double star with a relatively high-mass companion ($q=M_2/M_1=0.5$) and period $52.90\pm0.05\,d$. 4) Component F, is a CP star (B\,7.5\,p\,Si); its radial velocity, $23.2\pm4.2\,km\,s^{-1}$, is smaller than that of all other Trapezium members and, possibly, the evolutionary stage of the star is more advanced than that of members with similar mass. Consequently,  Component F is probably not physically related to the Trapezium. Several evidences point to the extreme youth of this stellar group; its further study, most likely, will shed light on the formation processes of massive stars.


\end{abstract}

\section{Introduction}

It has been a long time since I last participated in research on Gaseous Nebulae, so I apologize for my momentarily switching the subject matter of this meeting down to stellar affairs. 

In the last years I have been working on the spectral analysis of the Orion Trapezium stars, based on \'Echelle spectra ($R\simeq$ 15000) obtained with the 2.1-m telescope at the Observatorio Astron\'omico Nacional in San Pedro M\'artir, Baja California. Many colleagues have contributed to obtain the spectra, reduce the data and calculate orbital and eclipsing parameters, among which Juan Echevarr\'ia, Yilen G\'omez-Maqueo, Lester Fox and Alan Watson stand out. The aim of this work is to clarify some observational parameters of this famous and complex stellar group. 

\begin{figure}  
\begin{center}
\hspace{0.25cm}
\includegraphics[height=5.0cm]{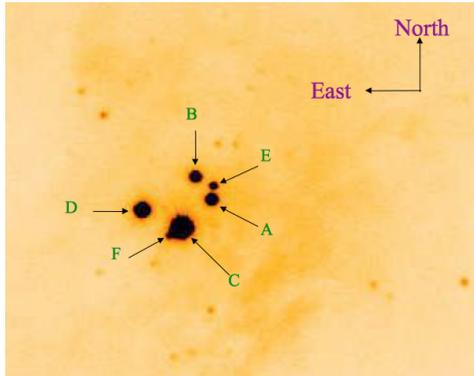}
\caption{ The Orion Trapezium in a two-second exposure image obtained with the $1.0\,m$ telescope at the Observatorio Astron\'omico Nacional in Tonantzintla, Puebla, using a Johnson $V$ filter. The primary mirror was covered except for two equal circular openings $20\,cm$ in diameter. Components A and D are separated by about $21\arcsec$. Notice that Component A was near minimum light during primary eclipse.}
\label{Trap1m}
\end{center}
\end{figure}

The Orion Trapezium ($\theta^{1}$\,Ori = ADS\,4186) is the center of the massive star forming region closest to the Sun and the main ionization source of the Orion Nebula. It consists of six bright stars (their $V$ magnitudes spanning from 5.1 to 10.3) that fit inside a circle $22\arcsec$ in diameter (see Figure~\ref{Trap1m}) and are immersed in the brightest part of the nebula.
These circumstances have undoubtedly contributed to the fact that there are many important observational properties associated with this notorious stellar system that are either uncertain or remained unknown until recently. Indeed, the Trapezium stars are difficult targets for small telescopes because of their mutual proximity, whereas they are too bright for most stellar surveys undertaken with bigger telescopes.
In particular, the radial velocity of most of its components is poorly known. This is understandable since, in addition to the already mentioned problems, nearly all of these stars are at least spectroscopic binaries. In fact, as we will show here, all but one of the six Trapezium stars are double-lined spectroscopic binaries, three of which are also known eclipsing binaries. It is important to obtain reliable velocity curves of these binaries, because the precise knowledge of their orbital parameters will allow a better calibration of the physical properties of high-mass young stars, in addition to understand their formation process and the evolution of massive multiple systems.
For instance, to illustrate the latter point, \cite{Allen+17} find that realistic N-body simulations of  the Orion Trapezium predict that the system will probably become dynamically unstable in a very short time ($< 30\,000\,y$) unless the mass of the components (as presently known) is substantially increased.

A few months before this meeting, a very high-resolution interferometric study of the Orion Trapezium Cluster was published \cite[]{GRAVITY+18}, where some of the spectroscopic binaries observed were spatially resolved and the existence of suspected companions were confirmed. These results highlight the importance of securing precise orbital parameters, derived from radial velocity curves, for the double and multiple members of this very young, compact and interesting stellar system. 

Here, a short review of what is known about the multiplicity and the orbital elements of the Orion Trapezium members is presented. In addition, some relevant (though preliminary) results of our ongoing research are given, including a brief discussion about the membership to this very young system of Component F, apparently the only single star in this very young and compact stellar group.

\section{Component A}
\label{a}

$\theta^1$\,Ori\,A=HD\,37020=V1016\,Ori is at least a triple stellar system, with one B\,0.5\ Main Sequence (MS) star and two intermediate-mass, pre-MS companions. Considering that the Orion Nebula Cluster (ONC) has been well studied for stellar variability since the XIX Century, it is surprising that the eclipsing \citep{Lohsen75} and spectroscopic \citep{Lohsen76,Bossi+89} binary nature of this $V= 6.73$\,mag star was discovered not long ago, rising speculations on a possible recent perturbation or capture in this system. In addition, $\theta^1$\,Ori\,A is a strong and variable radio \cite[GMR\,12 in][]{Garay+87} and X-ray ({\em COUP} 745) source. At its maximum, this object becomes the brightest radio source in de ONC. It is important to note that the radio source {\bf is not} the spectroscopic binary, as originally believed, but the third component in this system \cite[and references therein]{PetrMassi08}.

The light curve of V1016\,Ori during primary eclipse has been rather well established by \cite{Bondar+00} and by \cite{LloydStich99}. It is about one magnitude deep, only slightly color dependent in the $UBVI$ photometric range, and approximately 21 hours long. Though wide, its bottom is not flat and corresponds to a partial eclipse produced by a relatively opaque object passing in front of a similar-sized but much brighter star. The secondary eclipse has never been observed and is expected to be very shallow (just a few hundredths of magnitud in V). The period of the eclipsing binary, $P=65.433\,d$, obtained by both the above mentioned groups from the timing of primary minima, is frequently adopted when calculating the spectroscopic orbital parameters, as is the case of those determined by \cite{Vitri+98} and by \cite{StickLloyd00} using nearly equal archival data and a few additional measurements of their own. Not surprisingly, both groups reach nearly equal orbital parameters: $e = 0.65(3)$ $K_1 = 33(2)\,km\,s^{-1}$, $\gamma = 28(1)\,km\,s^{-1}$, $\omega = 180^{\circ}(4) $ (numbers in parenthesis are representative of the error in the last digit, as given by those authors). 

The physical properties of the secondary star of the eclipsing binary have eluded convincing identification, though the initial (and correct) educated guess by \cite{Lohsen75}, based on the eclipse light curve, has almost always been confirmed: it is a pre-MS star. \cite{VitriPlachinda01}, in a very high signal-to noise ratio (S/N) spectrum obtained during the descending branch of a primary eclipse, clearly detected low-excitation lines that they interpret to arise from the secondary and tertiary stars, the former with $T_{eff}\,\approx\,8000\,K$ and the latter with $T_{eff}\,\approx\,3500\,K$. Indeed, the average heliocentric radial velocity of the 13 lines attributed to the secondary (eclipsing) component is $128.8\pm5.5\,km\,s^{-1}$ and that of that of the 7 lines associated to the tertiary star is $33.4\pm11.6\,km\,s^{-1}$ (errors are the standard deviation from the mean). From the former radial velocity and the orbital elements obtained by \cite{Vitri+98}, the first reliable value of $q=M_2/M_1=0.19\pm0.01$ was obtained by \cite{VitriPlachinda01}. The radial velocity of the tertiary is, within errors, similar to the systemic velocity of the eclipsing binary. Although these results are very important, the spectral range these authors analyzed (5300-5365\,\AA) is small and the fit of the combined synthetic spectra with that of $\theta^1$\,Ori\,A is rather poor; however, their work is indicative of the excellent possibilities that high-dispersion spectroscopy may bring to the study of this stellar system.

Our early attempts (2004-2006) to improve the results for $\theta^1$\,Ori\,A obtained by \cite{VitriPlachinda01} were fruitful. In the high S/N spectra obtained inside and around three primary eclipses (at minimum light in two of them), we realized that the spectrum arising from the secondary star was detected, even outside eclipse, when doing a cross-correlation with them and that of an early G-type standard star, and that the width of the spectral lines from the primary star was definitively smaller when inside eclipse as compared with their width well outside eclipse. This change in line width during primary eclipse is due to the Rossiter-McLaughlin (RM) effect \citep{Rossiter24,McLaughlin24}, that consists of an anomaly in the observed radial velocity of the occulted component,  normally (when the the ecuatorial and orbital planes are coplanar) first rising above the orbital radial velocity, back to normal at mid-eclipse, and then below the expected radial velocity until the eclipse ends. However, in $\theta^1$\,Ori\,A, during the 21-hour long primary eclipse, the radial velocity of the primary star is always noticeably bellow the predicted orbital value, except perhaps in the first few hours. This abnormality is due to the highly inclined orbit of the secondary star with respect to the projected ecuatorial plane of the primary star (large orbital obliquity). Consequently, during the (partial) eclipse, the secondary occults mostly one side of the rotating primary, in this case the hemisphere receding from us. 

\begin{figure}  
\begin{center}
\includegraphics[angle=0,height=5.0cm]{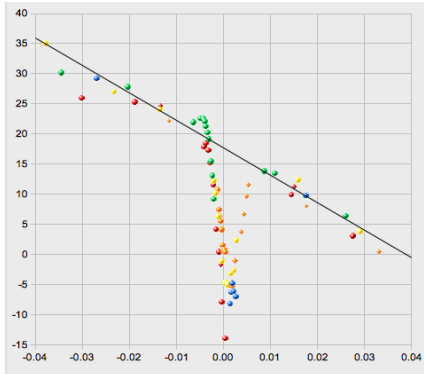}
\caption{Velocity curve of $\theta^1$\,Ori\,A obtained during and around seven primary eclipses, showing the abnormal Rossiter-McLaughlin effect. The horizontal axis is the photometric phase (see text for details). Colors and shapes of data points correspond to different observing seasons.}
\label{RMenA}
\end{center}
\end{figure}

The observed velocity curve inside and around primary eclipse, displaying the RM effect in $\theta^1$\,Ori\,A, is shown in Figure~\ref{RMenA}. In this figure, the horizontal axis is the photometric phase calculated using $P = 65.4330\,d$ and $HJD_\circ = 2\,453\,744.7585$. The latter date is that of the observed minimum light in the primary eclipse of 2006 Jan 6, calculated here from the photometric data obtained by Raul Michel Murillo and kindly made available to us. Note in this figure that the minimum radial velocity during the eclipse does not occur at phase zero (minimum light), but slightly later, and that there is a small anomalous rise in the velocity curve in the very inicial part of ingress. These features are indicative that the spin-orbit inclination is close to, but smaller than $90^{\circ}$. 

Other orbital parameters we derive for $\theta^1$\,Ori\,A (still subject to revision), using the above mentioned photometric period, are: $e=67\pm0.01$, $\omega=180^\circ\pm2^\circ$ and $q=M_2/M_1=0.20\pm0.1\,M_\odot$. They are in excellent agreement with those by \cite{Vitri+98} and by \cite{StickLloyd00}. From the mass ratio and assuming the mass of the primary to be $M_1=15\,M_\odot$ (that of a B\,0.5\,V star), the mass of the pre-MS secondary star is derived to be $M_2=3.0\,M_\odot$.

Preliminary analysis of the \'Echelle spectra obtained in 2004 and 2006, during and around three primary eclipses of $\theta^1$\,Ori\,A, was performed by \cite{Valle2011} for his BA thesis. All together, the spectra covered the complete ingress of the eclipse and well passed the minimum light. Adopting the projected rotational velocity, $v\,sin\,i = 55\,km\,s^{-1}$, obtained by \cite{SimonDiaz+06} of the primary component, and by means of a very simple numerical simulations (no limb darkening in the primary component; opaque, non emitting secondary) \cite{Valle2011} reproduced the observed RM effect by adjusting the ratio of the stellar radii ($r=R_2/R_1$) and the spin-orbit (obliquity) angle, $l$. In every simulation, the impact parameter (the minimum  projected distance between the two stars, in units of the primary star radius $R_1$, was fixed in order to fulfill one magnitude depth of the eclipse. The best fit found with this simple model occurs around $l=70^{\circ}\pm10^{\circ}$ and $r=0.8\pm0.1$. 

Such high orbital obliquity (spin-orbit angle) is totally abnormal in (eclipsing) binary stars. The only exception I know of is that of DI Her \citep{Albrecht+09,Albrecht+11}, where the equatorial plane of both members of the binary are strongly tilted with respect to their orbital plane; though no definitive explanation has been given to such misalignment, it could be the consequence of a relatively recent and strong orbital perturbation in a triple system, an extreme case of the Lidov-Kozai mechanism \citep{Lidov62,Kozai62}, or due to the capture of the secondary star by the primary. 
In any case, a third component in the system is required, either in a very eccentric orbit or ejected from the system as a consequence of the dynamical perturbation. Indeed, in $\theta^1$\,Ori\,A there is a third component (see below).
\cite{Valle2011} also derived the effective temperature of the secondary star to be $T_{eff}=5850\pm250\,K$, much lower than any of the other previous estimates. This was done by measuring  nine close pair line ratios sensitive to temperature, in two very high S/N spectra of the secondary star (obtained during minimum light in two eclipses), and comparing each of the same pair ratios in synthetic spectra created in the 5000-9000$K$ temperature range, with log g = 3.5 and solar abundances.

The third (hierarchical) companion in $\theta^1$\,Ori\,A, located about 0.2\arcsec north of the binary, was discovered by \cite{Petr+98} using holographic speckle interferometry in the $H$ and $K$ bands. In these bands, this star is about 1.4 magnitudes weaker than the out-of-eclipse, combined brightness of the binary components. As mentioned above, it is the highly variable (by at least a factor of 30) radio source that is frequently misidentified with the eclipsing binary; the physical process responsible for the radio emission and its large fluctuations has not been well established \citep[for details, see][]{PetrMassi08}. The possibility that this third component is an interloper was ruled out very recently, when the \cite{GRAVITY+18} proved that this interferometric companion is gravitationally bound to the eclipsing binary. When in cross-correlation of some of our high S/N spectra taken inside the primary eclipse, with a G5V standard, there is a hint of a late-type star at about the heliocentric systemic radial velocity of the eclipsing binary. This and other (currently preliminary) results await further analysis.

\section{Component B}
\label{b}

The weakest of the four stars that originally gave its name to the Orion Trapezium, $\theta^1$\,Ori\,B=HD\,37021=BM\,Ori, is at least a sextuple stellar system, so itself constitutes a subtrapezium or mini-cluster. Its brightest member is, as in the case of Component A, a hierarchical triple star also consisting of an eclipsing and spectroscopic binary (BM Ori, in a nearly circular orbit with period $P=6.4705\,d$) and a tertiary component, whose existence was recently, but inadvertently, confirmed beyond doubt by \cite{GRAVITY+18}. 
The reminding three components of the sextet are a resolved binary (separated from each other by about $0.12\arcsec$ and at nearly $0.97\arcsec$ from BM Ori), and a much fainter star located $0.6\arcsec$ northwest from the main component. More information about these companions may be found in \cite{GRAVITY+18} and references therein. The precarious dynamical stability of this subtrapezium has been studied by \cite{Allen+15}. In what follows, only the close and massive triple system will be analyzed.

The eclipsing binary, BM Ori, is a very peculiar Algol-type system: 1) The light curve of the primary eclipse seams to be a total one, with a wide flat bottom (though with small fluctuations) but, during this part of the eclipse, the spectrum of the star is almost equal to that outside eclipse; 2) the secondary eclipse is very shallow, only reasonably well observed in the R and I filters; 3) some time between 1990 and 2010, the duration of the flat bottom of the primary eclipse changed from about 8 hours to less than 4 hours \citep{Windemuth+13}. Several models have been proposed to explain the first two peculiarities, including the presence of opaque circumsecondary material, a strongly oblate Pre-Main-Sequence secondary, and a compact third star in the system \citep[see][and references therein]{PopperPlavec76,VasileiskiiVitri00}. The change in the duration of the eclipse was interpreted by \cite{Windemuth+13} as probably due to the very young secondary star actively accreting its circumstellar disk.

The main star in Component B is a fast rotator 
\citep[$v\,sin\,i = 240\,km\,s^{-1}$ according to][]{Abt+02} 
so it is a challenging spectroscopic target. On the visible range and at low dispersion, only neutral Hydrogen and Helium lines from the primary component are detectable, all of them contaminated by their nebular counterparts to a greater or lesser degree. The only exception is the Mg II $\lambda 4481 A$ line, that has a strongly variable, non-Gaussian profile. 
Consequently, it is not surprising that its spectrum has been assigned diverse classifications, from B0 to B4, and that some radial velocity data points and certain parameters of the binary orbit, derived from the primary component spectral lines, differ strongly between authors and epochs. 
The spectrum of the secondary star of this binary was first weakly detected by \cite{PopperPlavec76} in high dispersion photographic spectra, mainly during minimum light, as well as in the D, Na\, I doublet near quadrature. From the latter lines, these authors obtained orbital parameters for the secondary star and, from a group of He\,I lines in the photographic range, those of the primary component. These authors conclude that the secondary must be a late A or early F Pre-MS star with mass ratio $q = M_2/M_1 = 0.31$, where $M_1$ and $M_2$  are the masses of the primary and secondary stars, respectively.
 
It took some time for observers to realize that large (up to 30$\,km\,s^{-1}$) discrepancies in some radial velocity data, obtained at different epochs, were not only due to instrumental and measurement errors. \cite{VitriKloch04} first proposed that a third star was needed in order to explain the discordant orbital parameters and radial velocity outliers. 
They conclude that the center of mass of the eclipsing binary moves around the center of mass of the putative triple system in a very eccentric orbit (e=0.92) with $P=1302\,d$ and $K_{1,2}=20\,km\,s^{-1}$. The problem with this proposed orbit is that, at periastron (assuming both orbits are coplanar), the tertiary star lays inside the orbit of the eclipsing binary, as shown by \cite{Vitri+06}, who, additionally, cross-correlated a single high S/N spectrum of $\theta^1$\,Ori\,B with that of synthetic spectra calculated in the 5100-5500\,\AA~interval; in doing so, the latter mentioned authors clearly detect the secondary star with a $T_{eff} = 7000\,K$ synthetic spectrum, and claim to have detected the tertiary (with a much lower correlation height) with a $T_{eff} = 4000\,K$ template. Using their previously published orbital parameters \citep{VitriKloch04} and the spectroscopic mass ($M_1=6.3\pm0.3\,M_\odot$) for the primary star, they estimate of the masses of the two {\em satellite stars} to be $M_2=2.5\pm0.1\,M_\odot$ and $M_3=1.8\pm0.2\,$ where subscript 2 and 3 refer to the eclipsing companion and the tertiary star, respectively.
 
The work by \cite{Vitri+06} show that high resolution and S/N spectra of $\theta^1$\,Ori\,B, when cross-correlated with that of an early F-type narrow line star or with a $T_{eff} \approx 7000\,K$ synthetic spectrum, may yield precise radial velocities of the secondary component and, hence, accurate orbital parameters of the eclipsing binary, possibly with better precision than those obtained from the rapidly rotating primary star. In fact, the orbital eccentricity of the binary has been assumed to be zero in recent parameter determinations, even though in the first published orbital parameters, those by \cite{StruveTitus44} and \cite{Doremus70}, it was calculated to be 0.14 and 0.095, respectively. According to \cite{VitriKloch04}, the systemic velocity of the eclipsing binary, $\gamma_{1,2}$, is expected to vary with a period of about $1302\,d=3.56\,y$ due to the putative tertiary star; consequently, the orbital parameters of the close binary (derived from the secondary star) should be obtained from data secured during a single cicle or, at most, during very few consecutive orbital cicles. We did that in January 2010, when we acquired (at least) one high S/N spectrum of the star every night during nine consecutive nights. Setting $P=6.470524\,d$ \citep[the photometric period of BM Ori revised by][]{Vitri08}, we derive the following orbital parameters of the secondary star: $e=0.05$, $K_2=170\,km\,s^{-1}$, $\omega=82^\circ$ and $\gamma_{1,2}=5.9\,km\,s^{-1}$. Notice the very low value of the systemic velocity, as compared to $24\pm3\,km\,s^{-1}$ obtained in previous publications for this binary. This result alone is in agreement with \cite{VitriKloch04} proposal of a considerably massive tertiary. 

\begin{figure}  
\begin{center}
\includegraphics[angle=0,height=5.0cm]{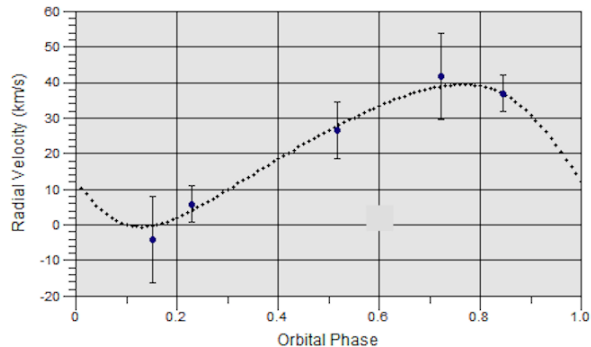}
\caption{Velocity curve of the center of mass of the eclipsing binary in $\theta^1$\,Ori\,B (BM Ori) folded by the orbital period, $P=1302\,d$, proposed by \cite{VitriKloch04} for the tertiary star. Error bars are estimates, mostly due to zero-point shifts yet to be determined.}
\label{B12Vrad}
\end{center}
\end{figure}

This result encouraged us to pursue further observations, ideally covering one cycle in each run, by exchanging observing time with other programs. With these data we derive $\gamma_{1,2}$ at different epochs from the velocity curve of the secondary star and by fixing all the other orbital parameters to those calculated in the January 2010 observing run (by far the best sampled one). In Figure~\ref{B12Vrad} the preliminary results of this work are shown; they correspond to the first nearly five years of observations (from 2010 Jan to 2014 Dec, spanning 1795 days). In this figure, the five available data points are folded with the 1302 d period proposed by \cite{VitriKloch04}; the vertical axis is the systemic velocity of the eclipsing binary system $\gamma_{1,2}$, obtained from the secondary star as described above. Care should be taken when interpreting this result since we are adjusting four orbital parameters of the triple system ($e, \omega, K_{1,2}$ and $\gamma$) with only five data points and, of course, there are other possible solutions, specially if the period is set free. What is clearly seen is that the systemic velocity of the eclipsing binary, $\gamma_{1,2}$, is indeed variable, with a semi-amplitude of about $20\,km\,s^{-1}$, in excellent agreement with \cite{VitriKloch04}, but with a much smaller eccentricity ($e \approx 0.3$ in the solution shown in Figure~\ref{B12Vrad}), surely a more stable configuration than that proposed by those authors. It is important to point out taht the stellar object detected by the \cite{GRAVITY+18} around the eclipsing binary in $\theta^1$\,Ori\,B is the same tertiary star \cite{VitriKloch04} postulate, and that its gravitational effects on the close binary are those shown here in Figure~\ref{B12Vrad}) ; at least, the one-year orbital segment obtained through the interferometric observations by the \cite{GRAVITY+18} is fully consistent with the 1302-day period.

\section{Components C and D}
\label{other}

Nothing can be added here to what is mentioned by the \cite{GRAVITY+18} about Component C. Summarizing, the brightest star in this subsystem, the intermediate O-type star that is the main source of ionizing photons in the Orion Nebula, is an oblique magnetic rotator, a rare characteristic that has been interpreted as evidence of a collision process in the formation of the star \citep{ZinneckerYorke07}. In addition, it is an interferometric binary, with a relatively massive companion (also detected in high S/N spectra), in an eccentric orbit ($e=0.69$) with period of 11.3 years. Additionally, it is suspected to be a spectroscopic binary, with a one solar mass secondary star in a 61.5.day period. The systemic radial velocity is, understandably, poorly known; probably its best estimate comes from the spectroscopic orbital parameters of the 11.3-year binary obtained by \cite{Balega+15}, derived from a few usable primary and secondary lines, from which $\gamma(C_{1,2})=29.4\pm0.6\,km\,s^{-1}$ was obtained.

 Component D ($\theta^1$\,Ori\,D = HD 37023), a $V=6.7\,mag$, spaectral type B\,0.5\,V and $T_{eff}=32\,000\,K$ according to \cite{SimonDiaz+06},  has been suspected to be a spectroscopic binary ever since its first spectral series was observed. However, efforts to find a period for it had been scarce, to put it mildly. \cite{Vitri02} gathered published radial velocities and measured available IUE spectra for this star; he derived two possible solutions for the orbit with periods $20.27 d$ and $40.53 d$. In some of our high dispersion and S/N spectra we clearly noticed that certain lines, namely the Si\,III triplet around $\lambda\,4560$ and C\,II $\lambda\,4267$, appear double when the weaker component is significantly blue-shifted. These lines reach their maximum intensity at around $T_{eff}=20\,000\,K$ (about type B\,2 in the MS), so it is reasonable to estimate that the secondary star is about two magnitudes weaker than the primary.

Considering $\theta^1$\,Ori\,D as a double-lined spectroscopic binary, we have measured radial velocities of this star in several high quality spectra obtained, randomly and whenever posible, during the last 14 years. From these we derive $P=52.90\pm0.05\,d$ and, from the primary component lines, the orbital parameters $e=0.42$, $K_1=36\,km\,s^{-1}$, $\omega=9.8^\circ$ and $\gamma=38\,km\,s^{-1}$. All these parameters are being updated and revised, specially the systemic velocity, that is surprisingly large when compared with the radial velocities of other Trapezium members. From those spectra in which the secondary star lines could be deblended from those of the primary, an average mass ratio $q=M_1/M_2=0.47\pm0.05$ has been derived, which is consistent with that expected for a binary made up of a B0\,V primary and a  B2\,V secondary.

The period and the eccentricity we obtain for $\theta^1$\,Ori\,D are equal, within errors, to those derived by the \cite{GRAVITY+18}, $P=53.0\pm0.7\,d$ and $e=0.43\pm0.7$, an amazing and wonderful result of modern interferometry for a spectroscopic binary at $400 pc$ from the Sun! A very weak additional visual companion, at $1.4\arcsec$ from the bright binary, has not been proved to be physically related to the binary \cite[for details see][]{GRAVITY+18}.

\section{Components E and F}
\label{other}

Components E and F are about 3.5 and 5.0 magnitudes weaker than, and located less than $5\arcsec$ to, their closer Trapezium members (Component A and Component C, respectively; see Figure~\ref{Trap1m}). This explains why their basic observational parameters were (and some, still are) poorly known.

$\theta^1$\,Ori\,E was discovered to be a a double-lined spectroscopic binary by Costero et al. (2006). Its nearly identical members are pre-MS stars with approximate spectral type G2IV; they are in circular orbit with $P=9.8952\,d$ and $K_1=K_2=84.4\,km\,s^{-1}$ \citep{Costero+08}. This binary is a variable radio \cite[e.g. see][]{Felli+93} and X-ray source source ({\em COUP} 732); it also varies by several tenths of magnitud in the visual range \citep{Wolf94} and by hundredths of magnitud in the $3,5\,\mu$ and $4.5\,\mu$ bands \citep{Morales+12}. However, the scarcity of published absolute photometry data on this star is amazing, except for its K-magnitude for which five measures are found in the literature, averaging about 6.8$\,mag$. In a {\em Spitzer} survey dedicated to the search and characterization of variability of the ONC stars, \cite{Morales+12} found that Component E is a grazing eclipsing binary and derived the mass of its members to be  $M_1=M_2=2.80\,M_{\odot}$ with 2\% accuracy. Together with its measured high- precision proper motion \citep{Dzib+17} and parallax \citep{Menten+07,Kounkel+17}, these results make of $\theta^1$\,Ori\,E the highest mass pre-MS star with well known physical parameters.

Component F has received very little attention and practically nothing was known about it. \cite{Herbig50}, in the last sentence of his early paper on the spectroscopy of variable stars in the ONC, just before the acknowledgements and after discussing the spectral type of Component E, writes: ``{\it Star F can be classified, with considerably more confidence as of type B\,8}''. This short entry was noticed and registered by \cite{Parenago54} in his catalogue of stars in the ONC, perpetuating the only spectral classification I know of, published for this star. We have obtained \'Echelle spectra of $\theta^1$\,Ori\,F in six nights spanning six years, in order to verify this classification and, of course, to find out its radial velocity and possible multiplicity. To our surprise, the metallic spectral lines in this star are very narrow and numerous. The spectral type we estimate from them is in excellent agreement with that given by \cite{Herbig50}, though in our spectra we register obvious peculiarities, like abnormally strong Si\,II, Si\,III and P\,II lines, indicating overabundance of these elements and the chemically peculiar (CP) character of the star (also called Ap stars). The He\,I lines, expected to be quite strong at $13\,000\,K$ (the temperature we estimate for it) are undetected, probably because they are both weak and filled-in by their nebular counterparts. Hence, we classify $\theta^1$\,Ori\,F as a CP B\,7.5\,p Si star. 

The mean heliocentric radial velocity we obtain for $\theta^1$\,Ori\,F from our six spectra is 
$23.2\pm4.2\,km\,s^{-1}$; 
the error is the standard deviation from the mean. The internal error in our spectra is expected to be about $2\,km\,s^{-1}$, so this result is not conclusive upon the star being a spectroscopic binary or not. Hence, together with the non detection of a companion to this star by the \cite{GRAVITY+18}, there is no convincing evidence of multiplicity in Component F.  Any way, this radial velocity is smaller than that of the three brighter Trapezium members (in particular, those of Components C and D).
It is interesting to note that the spectroscopic mass of this B\,7.5 star should be about $3.7 M_\odot$ if it is in the Main Sequence, slightly larger than that of the identical members of Component E ($2.8 M_\odot$) and of the secondaries of components A ($3.0 M_\odot$) and B ($2.5\,M_{\odot}$), all of them definitively pre-MS stars. So, if Component F is coeval with the other Trapezium members, it must be at or near the turn-on of the zero-age MS of the Trapezium Cluster. Alternatively, Component F is not a member af the Orion Trapezium, but an evolved star located by chance in front of theTrapezium, as proposed by \cite{Olivares+13}.

\section{Conclusions}
\label{discussion}

In agreement with the short-term dynamical instability that \cite{Allen+17} found in $\theta^1$\,Ori, the Orion Trapezium ($\theta^1$\,Ori) stellar system, we have shown here that there is clear evidence of extreme youth of the components of this nearby Trapezium: 

1) The large orbital obliquity (spin-orbit angle) we find in the eclipsing binary of Component A (part of a hierarchical triple system) is almost unique among binary stars and probably the consequence of tidal, secular friction with the tertiary member. 

2) The change of the eclipse duration in the close binary, also hierarchical triple Component B \citep[itself part of a six-member unstable mini-cluster,][]{Allen+17}, is probably the result of a sudden change in the circumstellar disk around the very young secondary star \citep{Windemuth+13}, possibly induced by the tertiary, for which we obtain a plausible, medium-eccentricity orbit. 

3) The oblique magnetic rotator nature of the hottest and most massive member of Component C, that possibly originated in a collision during its formation \citep{ZinneckerYorke07}, a process that could still be in progress through gravitational interactions between the massive interferometric companion and the relatively close, $1M_\odot$ spectroscopic companion proposed by \cite{Vitri02b} and \cite{Lehmann10+};

4) Component E ---a variable radio, infrared, optical and X-ray source--- is the highest-mass ($2.80 M_\odot$) pre-MS binary known, with most of the physical characteristics of its (practically identical) components determined with high precision.

Modern observational techniques will enable the precise dynamical study of the Orion Trapezium system that is probably in flagrant disintegration. They will also be important to better understand the physical characteristics that are still uncertain in several of its members, and to follow the evolution of those observables that are clearly varying. In doing so, I am sure, our knowledge of massive star formation will be greatly benefited.

\acknowledgments I am deeply grateful to the meeting organizers for this affectionate celebration, and for their including me in it. I am particularly grateful to Oli Dors for obtaining the generous support that made possible my participation. Finally, I sincerely thank Yilen G\'omez Maqueo Chew for her valuable suggestions that substantially improved this paper.

\bibliographystyle{aaabib}
\bibliography{Costero}

\end{document}